\documentclass{article}

\usepackage{PRIMEarxiv}

\usepackage[utf8]{inputenc} % allow utf-8 input
\usepackage[T1]{fontenc}    % use 8-bit T1 fonts
\usepackage{hyperref}       % hyperlinks
\usepackage{url}            % simple URL typesetting
\usepackage{booktabs}       % professional-quality tables
\usepackage{amsfonts}       % blackboard math symbols
\usepackage{nicefrac}       % compact symbols for 1/2, etc.
\usepackage{microtype}      % microtypography
\usepackage{lipsum}
\usepackage{fancyhdr}       % header
\usepackage{graphicx}       % graphics
\usepackage{siunitx}
\usepackage{natbib, url}
\usepackage{caption}
\title{Methods and measures for investigating microscale motility
}

\author{Karen Grace Bondoc-Naumovitz [1,$\ast$,$\dagger$], Hannah Laeverenz-Schlogelhofer [1,$\ast$,$\dagger$], Rebecca N. Poon [1], \\
\textbf{Alexander K. Boggon [1], Samuel A. Bentley [1], Dario Cortese [1], Kirsty Y. Wan [1,$\ast$]}\\
  (1) Living Systems Institute, University of Exeter, UK\\
  \texttt{$\ast$} Corresponding authors: k.bondoc-naumovitz@exeter.ac.uk;\\ 
  h.laeverenz-schlogelhofer@exeter.ac.uk; k.y.wan2@exeter.ac.uk\\ 
  \texttt{$\dagger$} These authors contributed equally.
}

\begin{document}
\sloppy

\maketitle

\begin{abstract}
Motility is an essential factor for an organism's survival and diversification. With the advent of novel single-cell technologies, analytical frameworks and theoretical methods, we can begin to probe the complex lives of microscopic motile organisms and answer the intertwining biological and physical questions of how these diverse lifeforms navigate their surroundings. Herein, we give an overview of different experimental, analytical, and mathematical methods used to study a suite of microscale motility mechanisms across different scales encompassing molecular-, individual- to population-level. We identify transferable techniques, pressing challenges, and future directions in the field. This review can serve as a starting point for researchers who are interested in exploring and quantifying the movements of organisms in the microscale world.
\end{abstract}

% keywords can be removed
\keywords{microscale motility \and organismal behavior \and biophysics \and microswimmers \and tracking}

\section{Introduction}
Motility is crucial in many aspects of life, enabling organisms to find resources, evade predators, and locate or colonize suitable habitats. By employing diverse molecular motor systems, an individual organism can convert chemical energy into mechanical energy and thereby control its movement \citep{Miyata2020, fletcher2004introduction}.
Swimming at the microscale is governed by fundamentally different fluid dynamics than swimming at the macroscopic length scale of our everyday experience \citep{Purcell1977LifeNumber}. A major difference between motion at the micro- and macro-scales is due to the relative sizes of the \textit{inertial} and \textit{viscous} forces, where inertia describes the tendency of an object in motion to remain in motion, and viscosity is the frictional force that slows down an object moving in a fluid. The ratio between these two forces is known as the Reynolds number ($Re$), where inertial or viscous effects dominate for high or low $Re$ respectively. For example, a human swimming in water has $Re \approx 10^6$, whereas a swimming \textit{E. coli} bacterium has $Re \approx 10^{-6}$. Microscopic organisms have evolved sophisticated self-propulsion mechanisms for navigating their highly viscous environment and aiding them in activities such as photosynthesis, feeding, or reproduction, which can increase their fitness or chances of survival \citep{Stocker2012EcologyOcean}.

Microbial communities are ubiquitous and underpin many biogeochemical cycles, meaning that the motility of microscopic organisms can influence food web dynamics and the structuring of ecosystems. The motility of photosynthetic (e.g. microalgae, diatoms, cyanobacteria), chemotrophic (e.g. archaea, bacteria), and heterotrophic (e.g. bacteria, ciliates, marine larvae) organisms can impact the flow of carbon and other nutrients in the food web and can affect small-scale spatial structuring of chemical and physical environmental factors \citep{Stocker2012EcologyOcean, fenchel2002microbial, fenchel1982ecology, worden2015rethinking, weisse2016functional}. With the advent of single-cell technologies and advancements in analytical and theoretical methods, we can begin to probe the complex lives of individual microscale organisms. However, bridging motility research across a continuum of physically and biologically relevant scales is daunting. Whereas population- and global-scale studies have shown how some organisms can shape large-scale ecosystem functioning through their influence on biogeochemical and nutrient cycling, studying the behavior of individuals can yield a more thorough understanding of their specific contributions. For example, \textit{in situ} observation of marine bacterial foraging reveals specific preferences toward chemical stimuli, which can define microscale partitioning of a community, as well as the remineralization rate of specific elements and nutrients in the ocean \citep{raina2022chemotaxis}. Meanwhile, several ciliated organisms, whether free-swimming (e.g. \textit{Paramecium}) or sessile (e.g. \textit{Vorticella}), forage by creating feeding currents through ciliary beating. Their ciliary arrangement can influence different feeding modes \citep{fenchel1982ecology, weisse2016functional} and predator evasion capabilities  \citep{nielsen2021foraging}. Their dual role as predator and prey influences the flow of carbon, and in turn the structuring of the trophic network in aquatic and terrestrial environments \citep{fenchel1982ecology, nielsen2021foraging, worden2015rethinking, weisse2016functional}. 

Measuring motility at the organismal scale is often challenging due to various technical constraints (e.g. broad range of relevant length scales, fast dynamics, and requirement for specialized and expensive equipment). Ultimately, we need to ensure that the methods (i.e. experiments, analyses, and models) are appropriate, reproducible, and practical and that the interpretation of the results is accurate, insightful, and can be meaningfully associated with the biology of the organism \citep{Berman2018}. Recent reviews have comprehensively described motility mechanisms grouping them taxonomically or based on their motility-enabling protein architectures \citep{Miyata2020, velho2021bank}. In contrast, here we focus on consolidating the different experimental, analytical, and mathematical methods used to study all microscale motility mechanisms across different scales from the molecular, to the individual and population levels. We thus identify commonalities between the various fields, which techniques could be transferable, and discuss common challenges and opportunities for future directions.

\section{Mechanisms of microscale motility}\label{mechanisms}

\begin{figure*}
    \centering
    \includegraphics[width=0.95\linewidth]{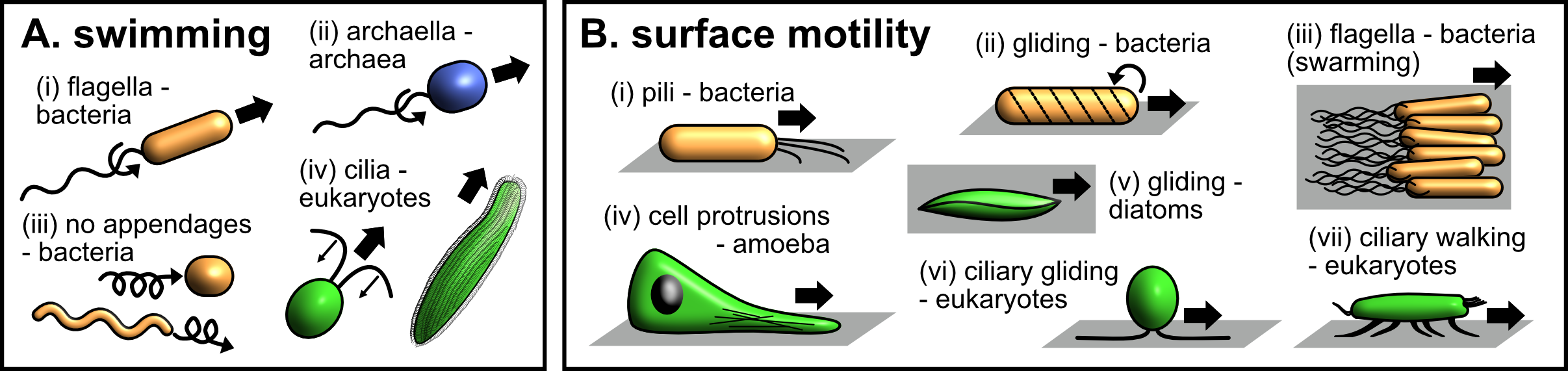}
    \caption{Schematic illustrations of the different microscale (A) swimming and (B) surface motility mechanisms.}
    \label{fig:mechanisms}
\end{figure*}

In this section, we summarize the mechanisms that microscopic organisms use to propel themselves through fluid or move across surfaces (as illustrated in Figure \ref{fig:mechanisms} and listed in Table \ref{tab1:mechanisms}). We note that the same style of locomotion (e.g. swimming, gliding, walking) can be achieved via different mechanisms and that the same motility apparatus can be used to achieve different types of movements. The diversity of mechanisms employed and locomotion behaviors performed by microscale organisms highlights the need to study microscale motility across different scales - from molecular mechanisms to the individual organism level and population scale.

\subsection{Swimming}
\textbf{Life at low Reynolds number.}
The Reynolds number, $Re$, is a dimensionless parameter that quantifies the ratio of inertial to viscous forces acting in a moving fluid. Microscale motility occurs at low $Re$, where the viscous forces experienced by such swimmers are much larger than the inertial ones. This imposes a variety of physical constraints on their motion. For example, when such a swimmer stops actively propelling itself, it will stop moving almost immediately. More subtly, such a swimmer can only propel itself by so-called `time irreversible' motions, in which a video of the swimming stroke looks different when played in reverse \citep{Purcell1977LifeNumber}. The side-to-side beating of a fishtail is not time irreversible, and such a swimming stroke at low $Re$ would not result in forward motion. In contrast, the `breaststroke' motion of the cilia of the low $Re$ swimmer \textit{Chlamydomonas} is time irreversible, leading to net forward motion.
An effective swimming strategy under such constraints is to exploit the large difference in the viscous drag coefficient experienced by a thin rod moving parallel or perpendicular to its long axis \citep{Becker2003SelfpropulsionReynolds}. Such drag-based propulsion via the use of long slender filaments is thus common across all domains of microscopic life. Yet despite this similarity, the propulsive machinery used by archaea, bacteria, and eukaryotes (namely archaella, flagella, and cilia, respectively) are entirely distinct, a striking example of convergent evolution \citep{Beeby2020PropulsiveCilia}.

\textbf{Flagella and archaella.} 
The bacterial flagella and archaella are both long, thin filaments (5-\SI{20}{\micro \m} in length, 10-\SI{30}{\nm} in diameter) driven by membrane-embedded rotary motors, with the former being more structurally complex than the latter. In bacteria, the rotary motor complex is powered by the ion motive force \citep{manson1977protonmotive, hirota1981flagellar} while in archaea a single ATPase is responsible for torque generation \citep{streif2008flagellar}.
In both cases, the torque translates into a helical waveform and the connected passive proteinaceous filament then acts as a propeller for cell propulsion. In bacteria, this torque transduction occurs via the flagellar hook while for archaea the filament connects directly to the motor.

\textbf{Cilia.} 
Despite appearing superficially similar, the structure of cilia is far more complex, comprising an order of magnitude more molecular components than either flagella or archaella \citep{Beeby2020PropulsiveCilia}. 
The axoneme section of the cilium produces the characteristic bending waves used for swimming. This structure typically consists of a central microtubule pair surrounded by a ring of microtubule doublets in motile cilia (Figure \ref{fig:methods}A.i) \citep{nicastro2006molecular}. 
Cilium bending occurs via the differential sliding of the outer doublets that is driven by dynein motors connecting neighboring microtubules \citep{satir1967morphological}. These dynein motors are regularly placed along the cilium allowing force actuation along its length. The cilium can thus produce far more complicated waveforms than those achieved with flagella or archaella.
Cilia are also widespread across multiple species of multicellular animals \citep{wan2020unity}. In particular, most bottom-dwelling marine invertebrate animals have a ciliated larval stage, where the cilia contribute to swimming, sensing, and feeding \citep{Marinkovic2020NeuronalFeeding}.  Cilia in marine larvae may be localized into bands as in \textit{Platynereis dumerilii}, or may densely cover the entire body, as in \textit{Nematostella} or in coral planulae larvae. In contrast to unicellular organisms, these multicellular ciliated swimmers can also modify their body shapes and trajectories by muscular action. Large numbers of cilia can also bundle together to form compound cilia that can propel larger organisms at higher $Re$, for example in ctenophores \citep{Jokura2022}.

\textbf{Swimming without appendages.} 
Swimming can also be achieved without the use of appendages, as in the bacteria \textit{Spiroplasma sp.} and \textit{Synechococcus sp.}. \textit{Spiroplasma sp.} lacks a peptidoglycan layer on its cell wall, rendering it flexible enough to change the helicity of its body to enable swimming by kink-propagation. The two ends of the cells have different handedness, and when the tapered end of the cell switches its helicity, a `kink' forms at the boundary of the axis which then propagates to the whole cell body enabling movement \citep{sasajima2021prospects, shaevitz2005spiroplasma}. Meanwhile, the swimming mechanism in \textit{Synechococcus sp.} is less understood. The current model is that it swims by forming small-amplitude waves through a helical rotor powered by proton-motive forces, similarly observed in the gliding mechanism model of mollicutes \citep{ehlers2012mysterious, brahamsha1999non}.

\begin{table*}
\caption{Overview of the main motility mechanisms discussed in this review. \label{tab1:mechanisms}}
\tabcolsep=0pt
\begin{tabular*}{\textwidth}{@{\extracolsep{\fill}}lll@{\extracolsep{\fill}}}
\toprule
mechanism of motility & domain(s) of life & example organisms\\
\midrule
flagella    & prokaryotes: bacteria     & \textit{Escherichia coli}, \textit{Vibrio alginolyticus} \\

archaella   & prokaryotes: archaea      & \textit{Halobacterium salinarum}\\

cilia       & eukaryotes: microalgae, dinoflagellates,   & \textit{Chlamydomonas reinhardtii}, \textit{Dinophysis acuta}, \\ 
    & ciliates, marine invertebrate larvae, rotifers & \textit{Paramecium caudatum}, \textit{Platynereis dumerilii}\\

swimming without  & prokaryotes: bacteria & \textit{Synechococcus sp.}, \textit{Spiroplasma citri}\\
appendages\\

pili        & prokaryotes: bacteria, archaea    & \textit{Pseudomonas aeruginosa}, \textit{Sulfolobus sp.}\\

gliding     & prokaryotes: bacteria    & \textit{Flavobacterium johnsoniae}\\
            & eukaryotes: diatoms     & \textit{Navicula sp., Bacillaria sp.}\\

cell protrusions    & eukaryotes: amoeba    & \textit{Dictyostelium sp., Physarum sp.}\\

\bottomrule
\end{tabular*}
\end{table*}

\subsection{Surface motility}

\textbf{Twitching.} 
As well as swimming in bulk fluid, many species of bacteria also move on surfaces. One form of surface motility is known as `twitching'. It is loosely defined as being an intermittent motion, such as that generated by the bacterial Type IV pilus, which repeatedly extends, adheres, and retracts to give a stop-and-go motion across a surface \citep{Burrows2012PseudomonasAction}. Much of this work has been carried out on the pathogenic bacterial species \textit{Pseudomonas aeruginosa}, but the Type IV pilus is also found in a wide range of bacterial and archaeal species, where it may be used for functions other than twitching motility. For example, in the soil bacterium, \textit{Myxococcus xanthus} twitching is utilized for social motility while single cells perform gliding (see section below) \citep{Mercier2020PilusMyxococcus}.

\textbf{Gliding.} Gliding can be defined as substrate-associated translocation of cells in the direction of their long axis without using any appendages such as cilia, flagella, or pili \citep{henrichsen1972bacterial}. This motility mechanism has been found in distinct lineages of eubacteria (e.g. cyanobacteria, myxobacteria, bacteroidetes, mollicutes), apicomplexans, and photosynthetic unicellular eukaryotes (e.g. diatoms) highlighting the convergent evolution of unique motility machinery in different organisms \citep{Miyata2020}. Gliding universally requires highly adhesive compounds -- commonly comprised of proteins and/or polysaccharides -- which are excreted onto the surface and are generally connected to the internal environment of the cell and onto the motor protein that provides the force required for movement. The distribution of these adhesive components over the cell surface can either be in a helical pattern as in the bacteroidetes \textit{Flavobacterium johnsoniae}, the myxobacteria \textit{Myxoccoxus xanthus}, and various filamentous cyanobacteria (e.g. \textit{Oscillatoria, Phormidium uncinatum, Lyngbya sp.}); or running in parallel to the long axis of the body for mollicutes, apicomplexans, and diatoms. Interestingly, only the photosynthetic microgliders (i.e. cyanobacteria and diatoms) continuously secrete a polysaccharide-rich slime-like substance as they move. Motors for movement are also highly diverse, including modified Type IV pilus-like complexes in filamentous cyanobacteria; rotary motors powered by the proton-motive force in bacteroidetes, myxobacteria and some mollicutes; and actin-myosin complexes in the eukaryotic microgliders (i.e. apicomplexans, diatoms). For an in-depth description of gliding mechanisms, see the following reviews for bacteroidetes, myxobacteria, and mollicutes \citep{mcbride2001bacterial, nan2014bacteria, nan2016novel, wadhwa2022bacterial}, for cyanobacteria \citep{hoiczyk2000gliding, wilde2015motility}, apicomplexans \citep{frenal2017gliding, heintzelman2006cellular}, and diatoms \citep{poulsen1999diatom, wetherbee1998minireview}.

\textbf{Protrusion-based locomotion.} 
Amoeboid movement is perhaps one of the first and most well-known of all the surface motility mechanisms, with most research focusing on the social amoeba and cellular slime mold \textit{Dictyostelium sp.} and the acellular and `many-headed' slime mold \textit{Physarum sp.}, as well as leukocytes. Organisms travel by changing their shape through protrusion and retraction of plasma membrane extensions (e.g. pseudopodia, blebs) and reversibly adhering to the surface \citep{lammermann2009mechanical, petrie2016multiple}. Force-generating modes for locomotion can be driven by actin-polymerization or hydrostatic pressure. In the former, polymerizing actin filaments can generate sufficient force to drive out membrane projections in the form of lamellipodia (flat, sheet-like branched actin filaments) or filopodia (long, thin needle-like actin projections). Meanwhile, hydrostatic pressure is formed due to actomyosin contractility. Myosin II activity triggers the formation of `blebs': localized protrusions formed by the flow of cytosol along a pressure gradient. Bleb retraction is regulated by F-actin and actin-binding proteins, together with myosin. In \textit{Dictyostelium sp.}, both these mechanisms are observed but the preferred mode is dependent on the prevailing level of myosin II activity, with higher activity correlating with bleb formation \citep{lammermann2009mechanical, raz2022blebs, petrie2016multiple}. In the case of \textit{Physarum sp.}, fan-like sheet protrusions (i.e. veins) are formed via cytoplasmic streaming governed by an actomyosin system. Fluid cytoplasm travels through the veins via propagating waves and is then converted to a more rigid version. This forms thin branch-like protrusions that can be used by the organism to explore its surroundings \citep{awad2022survey, oettmeier2017physarum}. Protrusions can also be used for moving in a 3D environment where adhesion to surfaces is not mandatory. When cells are confined in a 3D scaffold, the retrograde flow of actomyosin is sufficient to produce friction on the walls to propel movement \citep{petrie2016multiple}. Meanwhile, in a fluid environment, cells can form `side-bumps' or sideway protrusions in the rear of the cell which act like a paddle for swimming in the water, such as in \textit{Dictyostelium sp.} \citep{van2011amoeboid}.

\textbf{Cilia-based surface motility.}
Cilia are not only used for swimming. Some organisms, like \textit{Trichoplax adharens} \citep{Smith2015Trichoplax,Bull2021arxiv}, can use cilia to walk or crawl along surfaces. During walking, the cilia undergo a periodic stepping action, with a locomotor force generated while the cilium is in contact with the substrate. 
Walking motility is also observed in ciliates of the subclass \textit{hypotrichs}, which possess compound cilia called cirri on the lower surface of the cell, e.g. \textit{Euplotes} \citep{Larson2022,Lueken1996} and \textit{Stylonychia} \citep{Krause2010}. Each cirrus is comprised of bundles of cilia that act together as a single leg-like appendage.
Another form of ciliary-driven locomotion is a type of surface gliding, most extensively studied for the microalgal species \textit{Chlamydomonas} \citep{Bloodgood1988,Shih2013,Collingridge2013}. Unlike other cilia-based motility mechanisms, gliding does not rely on cilia bending movements, instead, it is powered by the intraflagellar transport mechanism, which results in longitudinal sliding movements of the ciliary membrane glycoproteins that enable the organism to move across solid surfaces \citep{Shih2013}. 
Several species of ciliated marine larvae also exhibit various surface motility behaviors controlled by ciliary and/or muscular action \citep{Santagata2008,Martin1978}.

\section{Current techniques for studying microscale motility}

In this section, we outline the main techniques available to explore the diversity of motility mechanisms at the microscale. Due to advances in high-resolution microscopy, high-speed imaging, micromanipulation, image segmentation and tracking, machine learning, and modeling low Reynolds number fluid mechanics, the techniques available to study microscale motility are expanding, and the possibilities that come with combining the cross-disciplinary approaches promise to broaden our understanding of microscopic life. Here, our focus is on the organismal scale, i.e. the experimental, analytical, and mathematical modeling approaches used to study the locomotion of individuals, but many of the techniques can be applied more broadly, e.g. to study population-level dynamics. Figure \ref{fig:methods} provides an overview of the experimental, analytical and modeling approaches discussed below.

\subsection{Experimental methods}

\textbf{Live imaging across scales.} 
Live imaging is the most direct approach for the experimental investigation of motility. Whether it is imaging the waveforms of cilia, obtaining trajectories of individuals or populations, or using particle image velocimetry (PIV) to reveal the fluid flow produced by a microswimmer, capturing videos of the dynamic behavior of motile organisms is the basis for building an understanding of the mechanisms of movement, their behavioral signatures and their response to stimuli. Most live imaging is limited in that it reduces 3D forms and trajectories into 2D. Techniques such as 3D tracking, micro-manipulation, and microfluidics enhance our ability to perform live imaging across scales but are particularly relevant at the organismal scale and are discussed in more detail below.

\begin{figure*}
    \centering
    \includegraphics[width=0.86\linewidth]{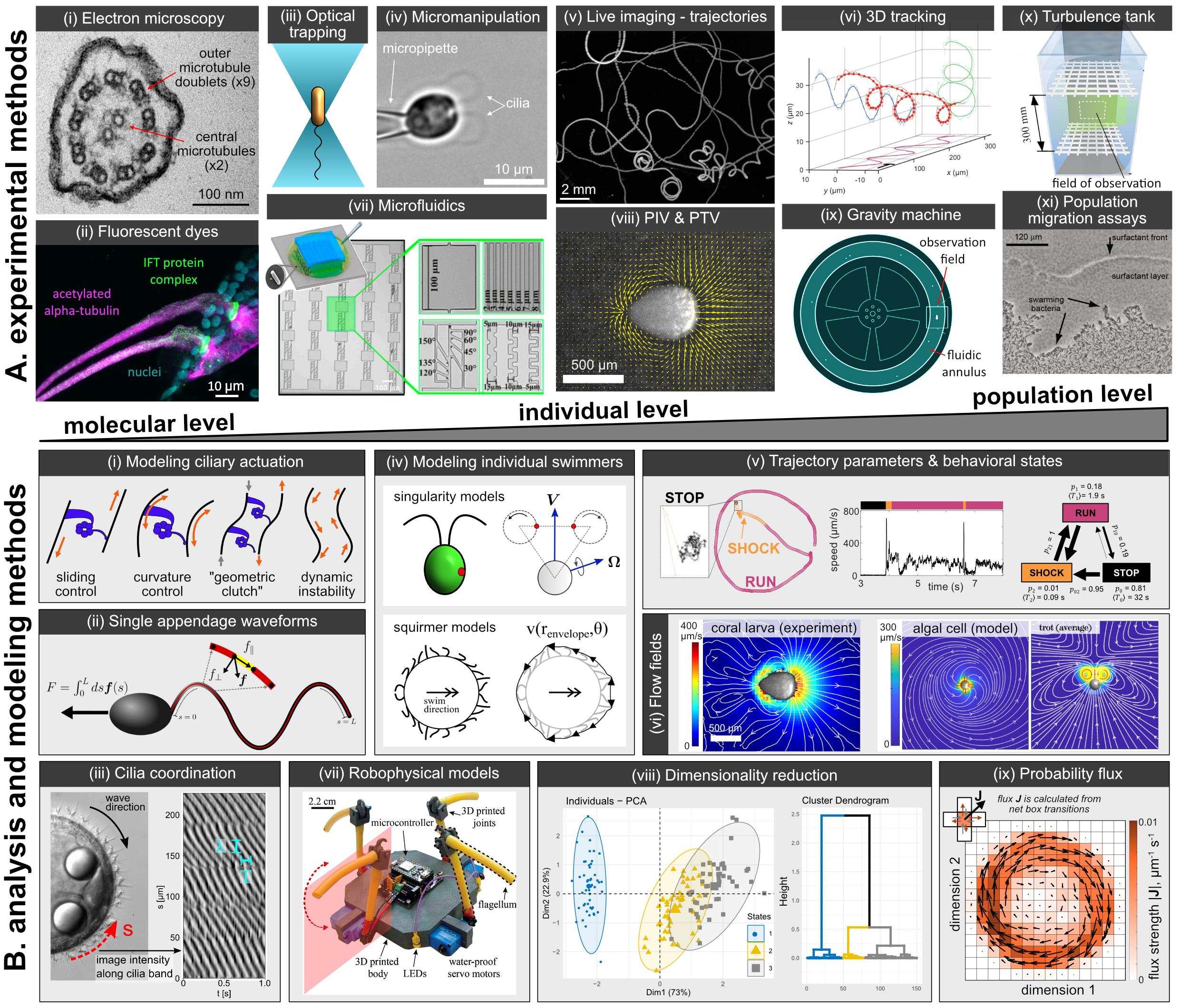}
    
    \caption{
    (A) Overview of the main experimental techniques organized by length-scale, from molecular to population- level.
    (i) Electron microscopy - a TEM image showing the cross-section of a sperm flagellum of the hydrozoan, \textit{Clytia hemisphaerica} (image credit: Kei Jokura).
    (ii) Fluorescent dyes - cilia of the ctenophore \textit{Bolinopsis mikadoin} immunostained to show acetylated alpha-tubulin (magenta), an intraflagella transport (IFT) complex protein (green) and the nuclei (cyan) (image credit: Kei Jokura).
    (iii) Optical tweezers - a schematic illustration of a bacterial cell held in an optical trap. 
    (iv) Micromanipulation - a \textit{Chlamydomonas} cell held by micropipette aspiration.
    (v) Live imaging, trajectories - a maximum intensity projection of jellyfish larvae trajectories recorded over a \SI{50}{s} time period.
    (vi) 3D tracking - a ciliate \textit{Tetrahymena} imaged via fluorescence microscopy forms right-handed helical tracks while swimming \citep{marumo2021three}.
    (vii) Microfluidics - a microfluidic chip design used to investigate bacterial swimming motility by observing their motion through geometries with various levels of complexity (adapted from \citep{Tokarova2021}).
    (viii) PIV \& PTV - the output from running PIV on a video of a coral larva swimming through fluid that has been seeded with passive tracer particles.
    (ix) Schematic of the `gravity machine' designed to track a single cell while allowing for free vertical movement (https://gravitymachine.org/).
    (x) Example of a turbulence tank with two horizontal grids that oscillate to generate turbulent flow \citep{Fagerstrom2022}.
    (xi) Population migration assays - the swarming behavior of a bacterial population can be observed via agar plate assays \citep{be2019statistical}.
    (B) Diagrams illustrating the main analytical and theoretical frameworks used to study microscale motility.
    (i) Modeling ciliary actuation - illustrations of the physical mechanisms underlying the four primary mathematical models for ciliary actuation.
    (ii) Single appendage waveforms - calculating the force vector per unit length on segments of a tracked cilium allows a prediction of the total force produced by the cilium.
    (iii) Cilia coordination - a frame showing the metachronal wave of the ciliary band of a \textit{Platynereis dumerilii} larva (left-hand panel). The pixel intensity along the ciliary band gives a proxy for the cilia beat phase, and can be plotted as a 2D function of distance $s$ along the band and time $t$ (right-hand panel). The period of the intensity oscillations in the $s$-direction gives the wavelength $\lambda$ and the period in the $t$ direction gives the ciliary beat period $T$. 
    (iv) Modeling individual swimmers - using singularity methods, the cilia beating of a \textit{Chlamydomonas} cell can be modeled by small beads constrained to rotate along circular orbits \citep{Cortese2021}. The squirmer model approximates the hydrodynamics of a densely ciliated swimmer by specifying the fluid velocity on an envelope that covers the tips of all of the cilia.
    (v) Trajectory parameters and behavioral states - the locomotor behavior of the octoflagellate \textit{Pyramimonas octopus} is classified into a trio of behavioral states based on the swimming speed. The state parameters (probabilities and expected durations) specify a unique reaction network \citep{Bentley2022}.
    (vi) Flow fields - flow fields for a coral larva, experimentally measured by PIV \citep{Poon2022CiliaryCoral}; and for an algal cell in top-down and sideways views, generated using a singularity model of the cell, and averaged over a whole beat cycle \citep{Cortese2022Gaits}. 
    (vii) A robophysical model of a quadriflagellate alga \citep{Diaz2021robophysicalmodel}.
    (viii) Dimensionality reduction - multivariate analyses can be used to reduce complex data by grouping the data in a low dimensional space through PCA and/or clustering procedures to tease apart the behavior of different organisms or assign behavioral states. 
    (ix) Probability flux - the probability flux strength for trajectories of single \textit{Chlamydomonas} cells trapped inside \SI{40}{\micro\meter} diameter microfluidic droplets indicates a preferred circling direction (adapted from \cite{Bentley2022}).
    }
    \label{fig:methods}
\end{figure*}

In larger-scale studies, measuring behavior in populations, especially \textit{in situ}, can be challenging due to  environmental factors that cannot be controlled as carefully as in the lab (e.g. light, temperature, nutrients) and the need for specialized equipment. However, most live-imaging techniques used on the organismal scale can be easily transferable to population-scale experiments in a lab setting.
The most classic experimental set-ups are capillary assays \citep{adler1966chemotaxis} or agar plates and other porous media \citep{nossal1972growth, be2019statistical} coupled with microscopy to track dynamic cell behavior \citep{Berg1972, bhattacharjee2019bacterial, taute2015high}. Meanwhile, Couette cylinders, turbulence tanks, or the newly developed gravity machine (Figure \ref{fig:methods}A.ix-x) can be coupled with PIV and microscopy to study the effect of laminar shear and turbulence on swimming organisms or sinking particles \citep{durham2013turbulence,arnott2021artificially, krishnamurthy2020scale}. Microfluidic devices can also be used to observe the dynamic behavior of cell populations (see section below).

\textbf{3D imaging.}
Conventional 2D imaging techniques are limited in their ability to fully resolve an organism's movement patterns, due to the fact that they can change their in-focus distance (i.e. z-position) while they swim. Therefore, tracking in 3D  provides unprecedented information on the motile behavior of microorganisms.
A benchmark study by Berg and Brown used a tracking microscope where the sample stage moves to maintain focus on a single \textit{E. coli} to determine motility changes in response to various stimuli \citep{Berg1972, berg1971track}. In recent years, various imaging methods (i.e. dual camera set-ups, fluorescence-based, defocused phase-contrast, and digital holographic microscopy) have been developed and improved to simultaneously track multiple cells in a 3D observation field \citep{bhattacharjee2019bacterial, taute2015high, wu2006collective, marumo2021three}.

By using two cameras to image swimming trajectories from different orientations, the two 2D images obtained can be combined to give 3D tracks of the organism. For example, this approach was used to study the phototactic response of the microalgae \textit{Chlamydomonas} and \textit{Volvox} \citep{Drescher2009}.

Fluorescence imaging relies on cells carrying fluorescent signals either by molecular labeling of cells, ingestion of fluorescent particles, or autofluorescence. Fluorescence imaging can be used to target specific features and improves signal-to-noise ratio, and therefore offers a range of approaches for 3D tracking. Fluorescently-labeled cells can be tracked in 3D using confocal microscopy \citep{bhattacharjee2019bacterial}, or by applying the tracking microscope approach to keep the individual in focus \citep{figueroa20203d}. By introducing additional optical components into a conventional epifluorescence microscope, and taking advantage of the point-like nature of fluorescent particles, \cite{marumo2021three} resolve the helical motion of the ciliate \textit{Tetrahymena} in 3D (Figure \ref{fig:methods}A.vi), by splitting the standard 2D image into two images such that the z-displacements of an object are transformed into the relative x-displacements of the split images. However, as fluorescence relies on signal intensity, it is limited in its spatiotemporal resolution.

Meanwhile, phase-contrast microscopy is especially useful for tracking transparent or colorless cells. When light passes through a cell, small phase shifts are translated into changes in amplitude, enhancing the contrast in the output image. Defocused phase-contrast imaging is a variant of this technique, where the z-position is inferred from the out-of-focus diffraction pattern, enabling 3D tracking using a conventional phase-contrast microscope \citep{wu2006collective, taute2015high}. 

In recent years, the use of digital holographic microscopy (DHM) for tracking cells has also been gaining traction. As the name implies, a hologram is constructed from the interference pattern between a light beam collected from the sample and a reference beam, both of which are split from a single laser beam. The resulting image contains the sample's phase and amplitude information, allowing detailed reconstruction of the 3D image. DHM comes in different set-up configurations but always consists of a light source, an interferometer, a camera (normally a CCD), and a computer. Applications of DHM range from tracking particles or free-swimming cells to flow fields, and from the lab to \textit{in situ} environments (for reviews see \cite{memmolo2015recent, yu2014review, garcia2006digital}).

\textbf{Micromanipulation.}
Free swimming individuals are often challenging to image at high magnification over long time periods. In order to observe the detailed waveforms of motile appendages and study long-term behavioral characteristics, the organism's body can be held fixed by micropipette aspiration (Figure \ref{fig:methods}A.iv) (e.g. see \cite{RufferNultsch1990, Brumley2014FlagellarInteractions, wan2014LagSlip}). Micropipettes are typically fabricated from glass capillaries using a micropipette puller. The inner and outer diameters of the micropipette must be carefully chosen such that it creates the necessary suction force while not sucking the individual too far into the pipette. Fire polishing the tip helps to create rounded edges to minimize the risk of damaging the organism \citep{oesterle2018pipette}. 

Micromanipulation tools can also be used to study how microscale organisms respond to stimuli. For example, a small glass stylus or microneedle can be used to apply a mechanical stimulus at a precise location \citep{Krause2010,Ogura1980}, micropipettes can introduce a localized flow \citep{wan2014rhythmicity} and cells held on a micropipette can be exposed to different controlled flow environments by holding them inside a microfluidic channel \citep{Klindt2016}.

When properly calibrated, micropipettes can also be used as force sensors by measuring the pipette deflections at high spatial and temporal resolution \citep{Schulman2014, Boeddeker2020}. For example, the forces produced by the beating cilia of \textit{Chlamydomonas} were measured by aspirating a cell to the end of a highly flexible double-L-shaped micropipette, which acts as a calibrated dynamic force cantilever \citep{Boeddeker2020}.

Micromanipulation techniques also enable electrophysiological experiments of microswimmers, for example, to investigate the bioelectric control of the beat direction, waveform, and frequency of motile cilia. This typically involves inserting a glass electrode into an individual cell and measuring its membrane potential, either to determine the organism's inherent electrical properties and spontaneous activity or reveal how the membrane potential responds to stimuli (e.g. current injection or mechanical stimulation). Electrophysiological experiments have been most extensively applied to study the ion channel properties and bioelectric control of ciliary beating in \textit{Paramecium} \citep{Brette2021}, but have also been performed with other ciliate species (e.g. \cite{Echevarria2016, Lueken1996, Hennessey2000, Krause2010}) and microalgae \citep{Harz1991}. These studies demonstrate the importance of the membrane potential in controlling motility and show that ions such as Ca$^{2+}$, K$^+$ and Na$^{2+}$ play a central role in controlling the movements of motile appendages and coordinating an organism's response to environmental stimuli.

Bioelectric signaling can also be studied by imaging the dynamics of calcium and voltage-sensitive dyes using fluorescence microscopy \citep{Xu2017,Grienberger2012}. While microelectrode recordings are typically more accurate and can achieve a higher time resolution, fluorescence imaging minimally disrupts an organism's behavior and does not require it to be immobilized. Fluorescent indicators of bioelectric activity can be genetically encoded \citep{Randel2014,Xu2017}, or introduced into the organism by incubating it with the relevant dye \citep{Alvarez2012}, biolistic loading \citep{Collingridge2013} or delivered via microinjection directly into the individual \citep{Iwadate2004}.

Microinjection is a technique in which a sharp micropipette is loaded with a chemical of interest and inserted into the organism for the intracellular delivery of fluorescent dyes or precise chemical stimuli. Microinjection is typically combined with microscopy in order to image the fluorescence signal and/or motility dynamics \citep{Tamm1994, Iwadate2004, Iwadate1997}. It has also been successfully performed in conjunction with electrophysiology experiments \citep{Nakaoka1990, Pernberg1995}. Microinjection techniques have been used to introduce calcium indicators into the cytoplasm and cilia to measure the calcium signaling dynamics associated with motility behaviors in, for example, the ciliates \textit{Paramecium} and \textit{Didinium} \citep{Iwadate2004, Iwadate1997, Pernberg1995}, and the ctenophore \textit{Mnemiopsis} \citep{Tamm1994}. It has also been used to control the intracellular concentrations of calcium and cyclic nucleotides to study their effect on ciliary beating in \textit{Paramecium} \citep{Saiki1975, Iwadate2008, Nakaoka1990}.

A different approach to micromanipulation is optical trapping (also known as optical tweezers), which uses highly focused laser light to generate optical forces able to manipulate objects that are typically nano- or micro-scale in size (Figure \ref{fig:methods}A.ii) \citep{Favre-Bulle2019}. Particularly relevant to the study of motility, optical trapping can be used to actively position and probe biological systems (including single molecules, organelles, and cells) \citep{Favre-Bulle2019, Ashkin1987}, which enables detailed observation of motility behaviors and dynamics \citep{Min2009}. Optical tweezers have also been used to measure the swimming forces generated by, for example, sperm \citep{Nascimento2008} and \textit{E. coli} \citep{Armstrong2020}.

\textbf{Microfluidics.}
Microfluidics involves the manipulation of fluids at volumes of micro-liters and smaller, using micron-sized channels. It has grown rapidly in recent decades due to its potent biochemical and medical applications, such as conducting immunoassays \citep{Weibel2005} or performing single-cell DNA barcoding on a large scale \citep{Zillionis2017}. It is also a flexible and powerful technique for studying motility at the microscale \citep{Son2015}.  

Microfluidic devices, which can be reproduced easily and consistently, can be designed with complex arrays of chambers or channels to create specific environments for motile microorganisms. There are numerous techniques to make microfluidic chips, but perhaps the most common is to use soft lithography to produce chips of polydimethylsiloxane (PDMS), which is ideal for research into living cells because it is non-toxic, gas-permeable, transparent, and relatively inexpensive  \citep{castillo2015microfluidics, RajM2020}. 
Chips can be designed to perform other functions such as mixing fluids \citep{Lee2011}, applying chemical gradients using permeable membranes \citep{DeJong2006}, or altering surface characteristics (e.g. hydrophobicity) via fabrication with particular chemical coatings \citep{RajM2020}.
In addition, droplet microfluidics can be used to confine cells further by trapping them in water-in-oil emulsions \citep{Bentley2022}. The ease of manufacture of chips allows for successive improvement of designs for rapid prototyping \citep{Zheng2012, Zheng2013, Zheng2014} or small design modifications to compare slight variations in environments \citep{Ostapenko2018}. 

 Microfluidic techniques have allowed the study of motility in such diverse microswimmers as bacteria, unicellular algae, and mammalian sperm cells. The natural local environments of microswimmers are heterogeneous; they can be open or highly confined \citep{Tokarova2021} and display complex boundaries and solid-fluid interfaces \citep{Thery2021}, such as the porous soil in which the motile microalga \textit{C. reinhardtii} lives or the mammalian oviduct that sperm cells swim through. Microfluidics is thus ideal for creating experimental environments that resemble the natural environments motile microorganisms must navigate.  
Both individual cells and large populations can be easily observed when placed within such a device, and in combination with microscopy and cell tracking, behavior can then be measured and analyzed using trajectory data for either individual cells \citep{Ostapenko2018, Bentley2022} or larger populations \citep{Kalinin2010, Rusconi2014}. This allows the observation of motility across spatial scales, which can give insights into the heterogeneity of behavior across the population and how individual organisms interact with their conspecifics. 
 
 One characteristic of microfluidic devices, which is key to their usefulness, is that at micron scales $Re$ is low and so fluid flow is laminar and thus (to some extent) predictable and easier to analyze \citep{Samuel2018, Schuster2003}. However, a drawback is that they are most useful for studying swimming organisms; studying other forms of motility such as surface-bound gliding motility requires careful consideration of the chemical and physical properties of the different surfaces (e.g. glass, PDMS) involved \citep{ducret2013single}. 

Unicellular swimming algae, primarily the model species \textit{C. reinhardtii}, are commonly used in microswimmer research due to the structural and functional similarity of their cilia to those present in higher mammals such as humans. Microfluidic devices have been used to study the interaction of swimming \textit{C. reinhardtii} and its cilia with surfaces \citep{Kantsler2013,Contino2015}, the effect of boundary curvature on cell location and concentration \citep{Ostapenko2018} and the cell response to light stimuli \citep{Bentley2022}.

Microfluidics has also been used in a range of experiments studying bacterial motility. It is a powerful tool for studying bacterial chemotaxis since consistent and reliable chemical gradients can be formed by fluid flow. For example, agarose gel can be used within microfluidic devices to produce a barrier to the fluid that allows the diffusion of small molecules across it, generating a consistent chemical gradient in an environment. These gradated environments have proven very useful for probing behavior and understanding the chemical pathways of tactic behavior, most extensively for \textit{E. coli} \citep{Ahmed2010,ColinandSourjik2017}.

Due to its ability to create highly controlled environments, microfluidics has been used in experiments to prove or test predictions made by multi-scale theoretical models \citep{Cammann2021,Kalinin2009,Tokarova2021}. For example, the work of \cite{Kalinin2009} confirmed that \textit{E. coli} has high sensitivity towards gradients of the chemoattractant amino acids, $\alpha$-methyl-DL-aspartate and L-Serine. Follow-up work \citep{Kalinin2010} demonstrated how \textit{E. coli} respond to multiple chemical gradients, common in natural environments but difficult to produce consistently in vitro, again demonstrating the utility of microfluidic-based methods. 

The mechanics of bacterial navigation and motility can also be readily studied in a microfluidic device. For example, the work of \cite{Tokarova2021} focused on the effect of high levels of confinement and boundary encounters in five distinct bacterial species (Figure \ref{fig:methods}A.vii). Again, microfluidics allows for the comparison of theoretical models with experimental motile behavior; in this case, \cite{Tokarova2021} compared experimental cell trajectories with models of bacterial wall interaction dependent upon cell size and flagellar arrangement. This work highlighted  the remarkable potential of microfluidics to reveal novel behaviors in microswimmers, such as helical motion in highly confined channels. The work of \cite{Binz2010} had a narrower focus but, in addition to observing higher cell velocities in \textit{S. marcescens} under confinement than in open field experiments, also demonstrated a similar zigzagging (or perhaps helical) behavior while in highly confined channels.

\textbf{PIV and PTV.}
Particle image velocimetry (PIV) and particle tracking velocimetry (PTV) are experimental methods for calculating the flow field of a fluid from high-speed video data (Figure \ref{fig:methods}A.viii).
In both methods, the fluid is seeded with passive tracer particles, and the flows are imaged at a high frame rate and resolution. The particle motion is measured from the videos and used to infer the fluid velocity as a function of space and time. Direct measurements of the flow fields around microswimmers can then be compared with those of other microswimmers and fitted to simple physical models. Since the inception of this idea almost thirty years ago, it has undergone many developments, particularly being revolutionized by the development of digital imaging \citep{Adrian2005TwentyPIV}. 

In (digital) PIV, the video frame is first divided into `boxes' or `windows' normally order $(10\times10)$ pixels square. Within each box, the 2D image correlation is then computed between one frame and the next. As the particles visible within the box will have moved with the fluid, the peak of the correlation gives the average velocity of the fluid within that box at that point in time \citep{Willert1991DPIV}. PIV produces velocity fields that are evenly sampled in time and space. It performs best when the particles can be homogeneously seeded throughout the flow, at a sufficiently high density. The spatial resolution is set by the analysis window size, which in turn is limited by the camera resolution, particle density, and fluid speed; the temporal resolution is limited by the camera frame rate. PIV naturally produces an Eulerian description of the flow, where the fluid velocity $\mathbf{u}$ is specified at fixed locations $\mathbf{x}$ (given by the box centers) at each time point $t$ (given by the successive video frames), giving a flow field $\mathbf{u}(\mathbf{x},t)$ as seen by a stationary observer who is external to the flow. 

PTV instead tracks individual particles as they move through the flow, allowing the fluid flow speed to be found at each position (in both time and space) that a particle passes through in the field of view. It is performed using a much lower density of tracer particles. The flow field produced by PTV is naturally a Lagrangian one, describing the fluid flow as experienced by an observer being carried along by it \citep{Virant19973DPTV}. The flow is specified by a function $\mathbf{X}(\mathbf{x}_0,t)$, giving the position $\mathbf{X}$ at time $t$ of a fluid parcel that started at position $\mathbf{x}_0$, thus describing how a parcel of fluid will move and deform over time due to the flow. 

The appropriate method depends on what kind of velocity information is required, and on the limitations of the experimental setup. The Eulerian flow field $\mathbf{u}(\mathbf{x},t)$ is generally more useful than the Lagrangian description for applications such as modeling individual microswimmers (see section below) \citep{Batchelor1967Introduction}. Whilst it is possible to convert between the two descriptions, if PIV is experimentally tractable then it may be simplest to perform PIV and access the Eulerian field directly. However, PTV can be preferable for certain experimental setups. For example, non-motile food particles used in feeding assays can also function as tracer particles, and a low particle density may be required for such feeding assays so that only PTV is possible \citep{Wandel2022ModulationTintinnid}.

In the basic setup, only the in-plane velocity can be measured, but setups such as scanning light-sheet microscopy \citep{Brucker1995DPIV}, holography \citep{pu2000advanced}, 3D PTV \citep{Virant19973DPTV}, and tomographic PIV \citep{Elsinga2006TomographicVelocimetry} can measure 3D flow velocities. 

A variety of parameters must be carefully chosen when setting up a PIV experiment \citep{Keane1990OptimizationVelocimeters, Melling1997TracerVelocimetry, Scharnowski2020PIVreview}. The particle size must be small enough to accurately follow the flow without altering it, while also large enough to produce clear images and so their Brownian motion is significantly slower than the motion of interest. In practice, the most commonly used sizes are between 1-\SI{10}{\micro\m}. The camera resolution and microscope magnification should then allow the tracer particles to be about 2-3 pixels in size in the images. The particle seeding density and analysis window size should be chosen together to give around 6 particles per box, and the particle displacement should be less than one-quarter of the linear box size per frame. In practice, a good choice of parameters for microscale swimming experiments might be: \SI{1}{\micro\m} diameter beads, camera resolution \SI{0.7}{\micro\m}/pixel, minimum window size 64 pixels square, particle volume fraction (density) $\approx 10^{-5}$ v/v, for a maximum particle velocity of 6 pixel/frame. For MATLAB users, the PIVlab toolbox is a user-friendly way of performing the analysis \citep{Thielicke2014PIVlab}.

\textbf{Molecular structure.}
To understand the motility mechanisms available to an organism, it can be informative to study the structure of the motility apparatus on a molecular level. Electron microscopy (EM) and confocal imaging can give detailed high-resolution insight into such structures (for details on applying such procedures to the model organism \textit{Paramecium} see \cite{AubussonFleury2015}). An organism's motility is determined both by \emph{what} structures the organism possesses, and \emph{how} they are used. For example, the maximum speed of a multiciliated organism will depend both on the density of the cilia and the frequency at which the cilia beat. Due to limited image resolution and complications due to fast-beating cilia, it is often difficult to measure cilia spacing by live imaging, and specimens must be fixed and imaged, normally by EM, to obtain such structural information. It is  possible to fix samples for EM instantaneously, giving a `snapshot' of the cilia behavior during normal swimming \citep{Larsen1991}. Electron microscopy, particularly TEM and cryo-EM, has also helped to reveal the internal molecular structures of motile appendages (Figure \ref{fig:methods}A.i). Such studies have been instrumental in showing that the locomotor force is generated along the whole length of a cilium, whereas for flagella and archaella the force is generated by molecular motors at the base \citep{Beeby2020PropulsiveCilia,wadhwa2022bacterial}.

Additionally, various structures in a specimen can be stained via appropriate antibody preparations and visualized using confocal microscopy (Figure \ref{fig:methods}A.ii). For example, immunostaining revealed the role of striated fibers in promoting basal body connections in the ciliate \textit{Tetrahymena} \citep{Soh2019} and cilia rootlets can be stained to show their preferred beating direction \citep{bengueddach2017BasalParamecium}. 
Recently, new sample preparation methods have led to the development of `expansion microscopy', which enables nanoscale resolution imaging with standard fluorescence microscopy by physically expanding fluorescently labeled fixed samples \citep{Wassie2019expansion,Gambarotto2019UExM}.
Most fluorescent imaging is limited in that it requires fixed samples, however, live imaging of the cytoskeleton can be achieved with specific fluorescent probes that stain the relevant protein filaments (e.g. tubulin or actin) \citep{Lukinavicius2014}.

\subsection{Analysis and modeling methods}

We now present an overview of analysis and modeling procedures (Figure \ref{fig:methods}B) that can be used to understand the vast experimental data sets produced by the different methods described above.

\textbf{Trajectory analysis.}
Videos provide observational evidence of how microscopic organisms move and how they respond to environmental stimuli. When observing an organism's movements, we are often faced with the questions - how can this be quantified? What are the meaningful parameters that describe its motion? What is the best representation of the organism's behaviors? How can quantitative techniques enrich our understanding? Can they reveal hidden dynamics not immediately obvious from observation alone?

Once images are acquired, the first step towards quantifying the motility usually involves some form of image segmentation, detection, and tracking in order to obtain trajectories. Various algorithms have been developed for these processes and a variety of commercial and open-source image processing and tracking programs are available (e.g. Imaris, Icy, Livecyte, TrackMate, TrakEM2 CellTrack, CellMissy, etc.), commonly automated for high throughput processing with multiple user interfaces for different use cases. For more information on algorithms and available tracking software, see \citep{boquet2021bioimage, chenouard2014objective, emami2021computerized, meijering2012methods, ulman2017objective}. 
Depending on the purpose of the study and the organism, tracking algorithms can follow the centroid position, the organism's shape, or the appendage waveform.
Among the available image processing platforms, the open-source program ImageJ is extensively used. In particular, its TrackMate toolkit provides effective feature extraction, segmentation, and tracking algorithms for obtaining trajectories, as well as some calculated motility parameters \citep{tinevez2017trackmate, ershov2022trackmate}.

From trajectories, we can obtain a coarse-grained description of movement characteristics. The most common track parameters are speed, turning angle, angular velocity, path curvature, and location (spatial distribution). Other related characteristics such as mean square displacement, and persistence measures (i.e. linearity, confinement ratio, asphericity, displacement ratio, etc.) can also give information about the organism's behavior. The most commonly measured motility parameters are outlined in various resources \citep{meijering2012methods, svensson2018untangling}.

Calculated motility parameters can be used to define the baseline behavior of organisms as well as their response to different environmental stimuli \citep{Bentley2022, Berg1972, Echigoya2022}. Furthermore, they can also be used in designating behavioral states (see section below), or as training data for machine learning, which can aid in high-throughput analysis for phenotyping behavior of cell populations \citep{choi2021emerging}. Various frameworks have been developed in this regard, which use several multivariate analyses or regression procedures to simplify the motility space prior to clustering or classification (see section below).

%for 3D - 3DeeCellTracker, a deep learning-based pipeline for segmenting and tracking cells in 3D time lapse images

\textbf{Behavioral states.} 
From microscopy observations and trajectories, we often find that an organism's movements can be categorized into a small set of behavioral states, with each state associated with a characteristic or stereotyped mode of locomotion, analogous to the walk, trot, and gallop gaits of a horse. Each gait typically involves a different mode of actuation in the motility apparatus. This type of description has been famously applied to the movements of \textit{E. coli}, which can be categorized into two states - periods of straight swimming when the flagella are bundled together are called `runs', interspersed with active reorientations called `tumbles' that occur when flagella un-bundle \citep{wadhwa2022bacterial, Berg1972}. Other motility strategies described using the behavioral states approach include `run-reverse-flick' in the bacterium \textit{Vibrio alginolyticus} \citep{Son2013, wadhwa2022bacterial}, a eukaryotic version of `run-and-tumble' in \textit{Chlaydomonas reinhardtii} \citep{Polin2009,Bentley2022}, `run-stop-shock' in the microalga \textit{Pyramimonas octopus} \citep{Wan2018RunStopShock,Bentley2022}, `helical-spinning-polygonal' swimming in \textit{Euglena gracilis} \citep{Tsang2018}, `roaming-and-dwelling' in the ciliate \textit{Tetrahymena} \citep{Jordan2013} and the `droplet-cone-trumpet' states in the ciliate \textit{Stentor coeruleus} \citep{Echigoya2022}. 

Behavioral states are typically classified using trajectory parameters such as speed, acceleration, track curvature, or cell shape. This generally requires the researcher to first asses the movement characteristics of the particular species and identify a subset of characteristic gaits. Setting thresholds for relevant parameters is often a suitable baseline approach to classifying states. Alternatively, clustering and other dimensionality reduction techniques (see section below) have also been used to identify behavioral states \citep{Echigoya2022,Larson2022}, which minimizes any potential researcher bias. With advances in machine learning it may soon be possible to find more unsupervised methods for identifying a discrete number of states from trajectories of any given organism, without the need to create custom algorithms \citep{choi2021emerging}.

By analyzing motility through the lens of behavioral states we obtain a low-dimensional description that can be useful in comparing the different strategies microscopic organisms employ to effectively navigate their surroundings. Once a trajectory is decomposed into a series of states, as well as characterizing the properties of the different states, we can use the discrete time series to specify a network, analogous to a chemical reaction network, with state probabilities, expected state durations and transition rates between the different states (Figure \ref{fig:methods}B.v). This discrete state representation allows us to quantify how sub-cellular dynamics change over time or in response to environmental cues \citep{Wan2018RunStopShock,Bentley2022,Echigoya2022}. 
%\textcolor{red}{add mention of markov models?}

\textbf{Modeling individual microswimmers.} Various mathematical descriptions have been investigated for a wide range of microswimmers. One main benefit of using a simple mathematical model to describe a swimmer is that it can be used to test the behavior and response of a swimmer to environments and conditions that may be difficult or impossible to create experimentally. Additionally, it gives insight as to the most important features of the organism in terms of its swimming -- if certain aspects of the real swimmer are irrelevant to the model, then that implies they are irrelevant to the swimming mechanism. Finally, it allows comparisons with the swimming of other organisms that can be described by the same model. Here we consider two main classes of model - singularity methods, and the squirmer model (Figure \ref{fig:methods}B.iv). Computational fluid dynamics can be used to implement more detailed and realistic swimmer models \citep{scherr2015CFDmicroorganisms}, but that is beyond the scope of this review.

\textit{Singularity methods}. In the $Re=0$ regime typical of microswimming, the Navier-Stokes equations governing the velocity of an incompressible Newtonian fluid reduces to the so-called Stokes equations \citep{Lauga2009TheMicroorganisms}. Singularity models are approximate solutions of the Stokes equation in the presence of a source of disturbance, such as a microswimmer in a given geometrical configuration or underlying flow. 

Singularity models are a form of multipole expansions, analogous to those used in electromagnetic and gravitational physics to express the fields at points distant from the sources. In the case of microswimmers, the source of the fluid velocity field is the swimmer, whose large-scale effect on the fluid can be approximated by the superposition of terms that correspond to different configurations of point sources. Examples include the field generated by a single point-force or monopole - called a stokeslet; that generated by a force dipole - which can be axisymmetric or composed of a symmetric (stresslet) and antisymmetric (rotlet) part; a quadrupole, and so on. These solutions are called singularities because the velocity field tends to infinity at the exact location of the source. Whether a term in this series is present in a specific model depends on the symmetries of the microswimmer and the obstacles (e.g. walls, other swimmers) surrounding it \citep{Blake1974SingularitiesBoundary}. Although the simplest singularity solution to the Stokes equation is the stokeslet, the vast majority of microswimmers are best modeled by solutions with zero net force, such as dipoles \citep{pedley1990ModelMicroorganisms}. The reason for this is that the swimming mechanism is typically due to internal forces, rather than external ones. Most swimming microorganisms are also not subject to external torques, which limits the mathematical expression of the dipole term to its symmetric part - the stresslet. The flow induced by swimming bacteria such as \textit{E.coli} is indeed well described by a singularity model including only a stresslet term \citep{Drescher2011FluidScattering}. 

The two point-forces composing a stresslet can either point toward the interior of the swimmer or toward the surrounding fluid. In the first case, the microswimmer is called a `puller' and propels by using its appendages or shape to pull the surrounding fluid toward its own body and redirect it sideways. On the other hand, `pusher' swimmers push the surrounding fluid outwards, thus swimming body-first. Typical pullers include \textit{Chlamydomonas} and many flagellated algae, while typical pushers are bacteria such as \textit{E.coli}.
Singularity models have been very successful at modeling the flows induced by cells swimming in a boundless fluid or near obstacles \citep{Berke2008HydrodynamicSurfaces, Drescher2010DirectMicroorganisms}, and also for understanding more complex three-dimensional behaviors such as super-helical navigation and phototaxis \citep{Cortese2021}.

\textit{Squirmer model.} The squirmer model of Lighthill and Blake \citep{Lighthill1952Squirming,Blake1971Spherical} is used to model multiciliated swimmers, which have cilia densely covering a large proportion of their bodies. It would be very computationally expensive to simulate or solve a model that includes the detail of such a large number of individual cilia. The squirmer model approximates the individual cilia by a continuous, approximately spherical, `envelope' that covers the tips of all of the cilia. It is then more straightforward to solve the fluid equations to find the flow resulting from the movement of this envelope. However, it is important, and sometimes not trivial, to choose an appropriate shape and speed for the cilia envelope. This is informed by the length and beat pattern of the individual cilia, and also any features of their global coordination (see section below). The spherical colonial alga \textit{Volvox} has been extensively modeled as a spherical squirmer \citep{Pedley2016SphericalSquirmers}. The squirmer model has also been used to study the behavior of multilicated swimmers near boundaries \citep{ishimoto2013SquirmerBoundary}, to compare the efficiencies of different forms of metachronal coordination \citep{Blake1971Spherical}, and has been extended to non-spherical body shapes \citep{theers2016ModelingSpheroidal,Zantop2020SquirmerConfinement}.

\textbf{Modeling ciliary actuation.}
Delving further into the swimming mechanisms of individual cells, significant attention has been given to understanding the actuation of cilia. The propulsive machines underlying cilium movement are the hundreds of dynein motors working in concert to bend the axoneme structure \citep{satir1967morphological}. How dynein activity is regulated in order to set up regular beating patterns has been the subject of several modeling approaches (Figure \ref{fig:methods}B.i) with no single model gaining a consensus in the field.

Three main models focusing on individual dynein activity regulation have been proposed. 
In each model, the activity of the dyneins on one side of the axoneme causes it to deform. This deformation bends the axoneme, eventually causing the dyneins to deactivate. Consequently, the dynein motors on the opposing side of the axoneme activate and reverse the bend. 
%Each of these models relies on choosing a critical value of some parameter at which this deformation occurs. 
Each of these models relies on choosing some parameter that triggers this reversal upon reaching a critical value. % RP is this right? 
In the sliding control model \citep{murase1991excitable}, there is an elastic resistance of the microtubule doublets to dynein-driven sliding and subsequent bending. The eventual build-up of resistance causes the dynein motors to detach from the neighboring microtubule. Whereas, the curvature control model \citep{machin1958wave, brokaw1972computer, sartori2016dynamic} relies on the deactivation of dyneins at a threshold value of the curvature of the axoneme, typically understood to take effect with some time delay. Meanwhile, the `geometric clutch' model \citep{lindemann1994geometric} relies on the assumption that dyneins are more likely to bind when the inter-doublet spacing is below a critical distance, controlled by a transverse force between neighboring doublets.

Instead of treating axoneme bending as the result of an antagonistic relationship between opposing sets of dyneins, models focusing on dynamic instabilities in flexible filaments have garnered interest in recent years. In this model, dynein activity produces a force tangential to the microtubule doublets. In a static filament, this would result in it buckling, however, if this force continually acts along the axis of the filament then the `follower force' can produce oscillatory waveforms in model cilia without the need for individual dynein regulation \citep{woodhams2022generation}.

\textbf{Cilia tracking, waveform analysis, and modeling.} Full appendage tracking provides data for comparison with theoretical models, having been used to study the regulation of dynein motor actuation in the cilia \citep{sartori2016dynamic}. The extracted waveforms can also be used directly to predict the swimming behavior in a simulated microorganism \citep{gallagher2018meshfree}.

Beyond characterizing the dynamics of microorganism appendages,  waveform tracking can be used to elucidate the force generated by these appendages (Figure \ref{fig:methods}B.ii). This is relevant for simple models of how locomotion is achieved in single-cell organisms and provides an approximation for the forces expected from bottom-up models of ciliary or flagellar molecular propulsion \citep{johnson1979flagellar}.

The simplest theoretical framework in which to determine the force produced by an actuated filament at low $Re$ number is that of local drag theory or resistive force theory (RFT) \citep{gray1955propulsion}. In this approach, one models the filament as a series of straight rods that experience a uniform force per unit length when driven by some external force, which in the case of microswimmers will be the propulsive machinery of the cilia, flagella, or archaella. Based on this approximation the fluid flow due to a deforming filament is replaced by that of a line of stokeslets (the fluid flow due to a point force) of the appropriate strengths. Using this approach, one can obtain an analytical form for the force produced by a moving filament in terms of the motion of the individual rods into which the filament is separated. This formulation is then ideally suited to analyze a tracked appendage that is already separated into discrete elements by virtue of the tracking. As such, RFT is regularly used as a method by which to evaluate the propulsive force generated by flagellated and ciliated microorganisms \citep{friedrich2010high, gray1955propulsion, velho2021bank}.

RFT does however have several significant limitations. The theory does not account for long-range hydrodynamic interactions, end effects at the filament tip, or the interaction between the cell body and the filament when used to predict the swimming dynamics of microorganisms. To account for these, an alternative theory was developed, initially by Hancock, called slender body theory \citep{hancock1953self}.

Slender body theory (SBT) differs from RFT by taking into account the (decaying) effect on the flow at a given point along the filament from points at increasing distances along the filament. SBT has received multiple rigorous mathematical treatments (see references in \cite{Lauga2009TheMicroorganisms}) but a more physically intuitive description was given by Lighthill \citep{lighthill1976flagellar}. His description of SBT involves modeling the flow at a point, $s_0$, on the filament as a superposition of the flow from the `inner' and `outer' problems. In the inner problem, the filament in the region near $s_0$ is modeled as a combination of stokeslets and source dipoles. In the outer problem, the filament further from $s_0$ is treated again as a line of stokeslets, because the dipole flow field decays spatially much faster than that of a stokeslet. This model captures the essence of SBT and makes it clear that it incorporates more fully the impact of interactions between different parts of a filament.

It is worth noting that the improved accuracy of SBT does come with a computational cost. As such, one should identify whether RFT remains an appropriate modeling choice for the problem under consideration, see \citep{johnson1979flagellar, walker2019filament}.

\textbf{Cilia coordination/metachronal wave analysis.}
Multicilated organisms overwhelmingly display some degree of coordination in their ciliary beating. In order to quantify and analyze the coordination dynamics, it is necessary to extract the phase of the cilia from the video data. When the cilia are widely spaced, the individual cilia can be tracked as described above. Such analysis of \textit{Chlamydomonas} has shown that the dynamics of its two cilia are more complex than simple synchronous beating \citep{wan2014LagSlip}. 
Many organisms, from unicellular ciliates such as \textit{Paramecium}, the colonial alga \textit{Volvox}, through to larvae of marine invertebrates such as \textit{Platynereis}, have large numbers of cilia, distributed all over their body, or localized into ciliary bands. Such arrays of multiple cilia usually coordinate into metachronal waves, where cilia organize into synchronously beating rows, with a constant offset between the beat phase in each row, leading to a `Mexican wave' pattern of the beating.
In such multiciliated systems, the beat phase must normally be inferred from intensity fluctuations within carefully chosen windows, due to the high cilia density. For example, in a video of a beating ciliary array, periodic oscillations of the intensity over the array give a proxy for the beat phase at that point, (Figure \ref{fig:methods}B.iii) \citep{Wan2019ReorganisationCoeruleus}. Where the imaging resolution is insufficient to resolve the cilia waveforms, local variations in the fluid flow velocity can be used instead as a proxy for the beat phase \citep{Brumley2015MetachronalOrigin, Poon2022CiliaryCoral}. 
Video data can thus be analyzed to find parameters such as wavelength, frequency, direction, and coordination length- and time-scales \citep{ringers2023CiliaZebrafish}. Measuring such parameters in experimental systems allows comparison with metachronal wave models such as those of \cite{Meng2021ConditionsCilia} and \cite{Solovev2022SynchronizationDominates}, and can inform simulations of the cilia-driven swimming of organisms \citep{Blake1971Spherical, Ito2019SwimmingModel}. Finally, metachronal coordination is not limited to ciliated swimmers and can be observed in a broad range of organisms, for example in the ctenes of ctenophores and the legs of shrimp \citep{Byron2021MetachronalDirections}.

\textbf{Physical modeling.}
Microswimmers can also be modeled using macroscale physical models. The physics of the low $Re$ regime can be recovered at this larger scale by choosing a fluid of suitable density and viscosity to give a $Re$ comparable to that of a microswimmer in water. In a similar way to computational models, physical models use a `bottom-up' approach to study the system, by implementing the minimal number of components necessary to reproduce the basic swimming behavior of the organism. For example, a minimal robophysical model of a quadriflagellate swimmer (Figure \ref{fig:methods}B.vii) can successfully reproduce the relationship between gait and swimming performance observed in real microalgae \citep{Diaz2021robophysicalmodel}. Artificial ciliary arrays can be programmed to perform a metachronal wave and used to investigate the effect of different wave parameters on various properties of the fluid flow \citep{Dong2020BioinspiredCoordination}.

\textbf{Dimensionality reduction and clustering techniques.} 
The high-speed and long-term imaging required to capture dynamic motility behaviors often produces complex high-dimensional data sets, whereas locomotor strategies are often highly stereotyped and low-dimensional. This is a recognized challenge in neuroethological studies of animal behavior and recent advances in quantitative analysis frameworks and machine learning enable low-dimensional descriptions of organism behavior to be achieved \citep{Berman2018, Datta2019Neuroethology}. %Bialek PNAS 2022
Approaches used in animal behavior research can be usefully applied to study motility in microscopic organisms since in both cases the raw data often consists of movement trajectories or videos of the individual's body postures. 

%It can be useful to consider the search for a low-dimensional description of motility from both a bottom-up and top-down perspective. In a bottom-up approach, we seek to first understand the locomotor mechanisms and the different ways they can be deployed, then relate this mechanistic understanding to the possible movement patterns or gaits of the organism. In a top-down approach, data from trajectories (like speed, track curvature or body positioning) can be analysed using techniques such as principal component analysis (PCA) and clustering to define a small subset of possible gaits. 

Standard multivariate analyses can be powerful tools for understanding and visualizing the multi-dimensionality of large datasets produced by track analysis. Mapping the data in a lower-dimensional space through principal component analysis (PCA), t-SNE (t-distributed stochastic neighbor embedding), or UMAP (Uniform Manifold Approximation and Projection) reduces the complexity of datasets and removes the noise while preserving important characteristics of the original data. Clustering techniques such as hierarchical clustering on principal components (HCPC) and \textit{k}-means assess the robustness of the grouping, the results of which are commonly depicted as a dendrogram (Figure \ref{fig:methods}B.viii).

PCA is a procedure for decomposing a dataset into a series of orthogonal modes forming a coordinate system describing its variance. These modes can then be used to reconstruct the original data, and depending on the number of modes chosen, one can capture a defined amount of the total variance in the data. Mathematically, this can be approached by computing the singular value decomposition of the data in question \citep{brunton_kutz_2019} or by calculating the eigenvalues and eigenvectors of a covariance matrix constructed from the data \citep{werner2014shape}.
%\ab{PCA has been widely used in the animal behavior field for stereotyping locomotion strategies of organisms including worms, flies, and rodents \citep{Berman2018} highlighting its multi-scale applicability. - remove? Doesn't really say anything not clear from the paragraph two above in reference to Berman review.}
While PCA is a linear dimensionality-reduction technique that identifies the most important features while preserving variance in the data, both t-SNE and UMAP are non-linear techniques where the probability distribution of the data is mapped based on the similarity of each observation. t-SNE and UMAP use different algorithms for calculating similarities with the latter being significantly faster, scalable (i.e. can be used in larger data sets), and better at preserving the local structure of the data \citep{mcinnes2018umap, van2008visualizing}.
Despite the power of dimensionality-reduction techniques, applications in microscale motility studies are less common. However, examples include identifying the basic waveforms and fluid interactions that drive propulsion in sperm cells \citep{ishimoto2017coarse, ma2014active}, %saggiorato2017human
assessing the possible number of states/gaits of a moving organism \citep{werner2014shape, kimmel2018inferring}, and phenotyping motility of a population \citep{martinez2011statistical, heryanto2021integrated, xin2022time, schoenauer2015generic, kimmel2018inferring, Echigoya2022}.

After reducing the features and the noise in the data, its results can be used for subsequent clustering  analysis to refine further the groupings that can help in understanding patterns and trends in behavior \citep{martinez2011statistical, heryanto2021integrated, xin2022time, schoenauer2015generic}. HCPC, as the name implies, uses the extracted principal components and performs iterative partitioning through splitting and joining groups either via a top-down (divisive) or bottom-up (agglomerative) algorithm based on similarities. In the divisive algorithm, the whole data starts as one big cluster, while in the agglomerative, each data point starts as its own cluster.  Based on the number of clusters resolved by HCPC, one can subsequently use \textit{k}-means clustering, to refine the ideal number of clusters. \textit{k}-means can also be used independently of the dimensionality-reduction techniques and HCPC but depending on the dataset, could yield non-meaningful results.

\textbf{Probability flux.} 
To further investigate how motility mechanisms and stochastic behaviors are associated with low-dimensional characteristics within high-dimensional parameter spaces, the concept of probability flux from statistical physics provides a useful measure to account for the arrow of time, characterize non-equilibrium dynamics, and reveal hidden patterns in motility behaviors. Once a parameter space of interest has been identified, the probability flux provides a heading and a strength according to the most probable trajectory direction starting from the current position in the chosen parameter space. Probability flux analysis was first introduced by Battle et al. \citep{Battle2016} to study the period beating dynamics of an isolated beating cilium of \textit{C. reinhardtii} in a phase space representing the cilium shapes. The approach has since been applied to analyze the long-time trajectories of individual microswimmers in confined physical geometries, revealing the emergence of self-organized flux loops (Figure \ref{fig:methods}B.ix) \citep{Cammann2021, Bentley2022}.
%Probability flux analysis - dynamical systems, non-equilibrium dynamics. Time-averaging.

\section{Where are we going?}

In this review, we have highlighted the experimental, analytical and mathematical techniques one can use to quantitatively characterize microscale motility, allowing measurable descriptions of behavioral dynamics. This enables us to gain insights into how an organism performs in a dynamic environment by overcoming or even exploiting the constraints placed on it by the laws of physics \citep{Wan2021OriginsExcitability}, for example, whether motility is beneficial in a turbulent environment for a cell in a low $Re$ regime. Quantitative analysis of behavior is also beneficial when comparing different organisms and for using experimental data to test hypotheses.

\textbf{Technical challenges.}
In order to advance our understanding of microscale motility, several technical challenges remain. First, although many tracking programs are available, they usually require time-consuming optimization and customization steps in order to make them applicable to the specific organism of interest. Current tracking methods are most suited to round objects with high contrast. Automated tracking is especially difficult when the object of interest has a time-varying shape. Therefore, we need more general segmentation and tracking algorithms that are applicable to a wide range of morphologies and movement characteristics, while also being simple enough so that they do not require extensive coding experience. 
In the field of animal behavior, several machine-learning-based algorithms have been developed for tracking animal position and posture \citep{pereira2022sleap,lauer2022multi}. It would be beneficial to develop similar platforms for tracking the movements of microscopic organisms, including changes in shape and appendage actuation, which is applicable to the wide range of morphologies and does not rely on high contrast imaging of cells at low density. 
Machine learning is also emerging as a useful approach to extract meaningful information about spatiotemporal features of cellular motility from imaging data and its potential use in phenotyping motility behavior is an area that could be developed further \citep{choi2021emerging}.
Another major technical consideration is data management. When acquiring high-magnification and high-speed videos often required to record microscale motility dynamics, large volumes of data can be accumulated (e.g. an experimental study can generate terabytes of data). Therefore, when planning such experiments it is crucial to invest in cloud storage or hard drive data management solutions and to carefully plan the data processing pipeline. 

\textbf{Moving beyond model organisms.}
Most fundamental knowledge on organismal behavior is from studying model organisms (e.g. \textit{E. coli} for bacteria and \textit{C. reinhardtii} for microalgae), partly because they are amenable for genetic manipulation which allows testing of specific motility machinery or signaling processes related to behavior. However, model organisms are not representative of the range of behaviors possible for a given motility mechanism and so generalizing can be inaccurate. There is a need to diversify study organisms, which can give new information on how, for example, the morphology or ecology of an organism (i.e. its niche) modifies behavioral patterns. For microswimmers, a repository of swimming kinematics exists as a tool for comparing movement characteristics \citep{velho2021bank}. However, such a tool does not exist for surface-based mechanisms (e.g. gliding). With the recent development of genetic tools such as CRISPR technology, one can also create genetic mutants, and explore the questions that are traditionally only possible by using model systems. By doing so, we can look at a broad range of related species and assess the similarities arising from evolution but also the differences specific to that organism.

\textbf{Linking different scales.}
Perhaps one of the biggest challenges is to connect understanding across different length scales, i.e. from the mechanics of molecular motors and the locomotor behaviors of individuals to large-scale community processes and biogeochemical cycles. This challenge is not entirely new and has been a prevalent question in behavioral and migration studies of macro-organisms (e.g. insects, birds, whales). A recent framework that attempts to bridge this disconnect is movement ecology, which provides a way to link the physiological and behavioral properties of individuals to movement patterns across spatial and temporal scales. Movement ecology is based on four different factors - the movement mechanism, the internal state of the organism, the navigation and re-orientation capabilities, and the environmental context of the organism. This framework combines insights from cell biology, ecology and evolution, which has promising potential to synthesize a more thorough understanding of the causes and consequences of locomotion \citep{wisnoski2022scaling}. Additionally, quantitative analysis of experimental data combined with theoretical modeling is a powerful tool for bridging the gap between scales and building a cohesive understanding of behavior. As discussed throughout this review, modeling allows testing/simulating conditions that cannot be explored experimentally and experiments can be used to validate models. An example of this multi-scale approach is in investigating the dynamics of harmful algal blooms by integrating studies on molecular biology, individual and collective organismal behavior (e.g. gyrotaxis and vertical migration), and the physical environment (e.g. turbulence, nutrient availability) through various modeling approaches, in the hopes of improving prediction and forecasting \citep{berdalet2014understanding, franks2018recent}. 

\textbf{Collaborating across disciplines.} 
In order to successfully elucidate different aspects of microscale motility, specific skill sets and knowledge from varied disciplines need to be combined. 
Traditionally, the `why' questions of function and evolution might be viewed as the premise of biologists and ecologists, whilst physicists, mathematicians and engineers ask the `how' questions of forces and mechanics.
The methods of investigating behavior can vary between different fields, hence generated knowledge is specific to the scale and design of the study. 
Behavior has a multi-faceted nature, therefore the integration of different techniques and disciplines by working collaboratively can give rise to more thorough insights into the multi-scale aspects of behavior. 
Although interdisciplinary collaborations already exist, the difficulty lies in the lack of a shared foundation for what is considered common knowledge. Thus, there is a need to simplify communication to enhance the flow of information.  
An example of such a cross-disciplinary initiative is the `motile active matter roadmap' by \cite{gompper2020}, which brought together researchers from diverse disciplines to assess the current state of the art of the active matter field.

We hope this review can be a starting point and toolkit for researchers looking to describe behavior quantitatively in new and exciting systems. Here, we have highlighted both the limitations and the scope of what can actually be measured from experimental systems to test model predictions, while also identifying the areas where modeling would be particularly useful.
The field is ripe for researchers to conduct quantitative analysis, widen the diversity of study organisms and collaborate across disciplines in order to drive real progress in our understanding of the multiscale processes of microscale motility.

%%%%%%%%%%%%%%

\section*{Competing interests}
No competing interests are declared.

\section*{Author contributions statement}
K.G.B.N. and H.L.S. conceived the topic for this review. All authors contributed to writing and reviewing the manuscript.

\section*{Acknowledgments}
This work was funded by UK Research and Innovation (UKRI) under the UK government’s Horizon Europe funding guarantee [grant number EP/X02119X/1] (K.G.B.N), and the European Research Council (ERC) under the European Union's Horizon 2020 research and innovation program grant $853560$ EvoMotion (K.Y.W). K.G.B.N was also supported by the Integrative and Comparative Biology Society to attend and contribute to the associated symposium. We also acknowledge Kei Jokura for providing the EM and fluorescence images used in Figure \ref{fig:methods}A.

\bibliographystyle{abbrvnat}
\bibliography{references}

\end{document}